\documentclass[12pt,a4paper]{article}

\usepackage[dvips]{graphicx}
\usepackage{cite}
\usepackage{amsmath}
\usepackage{amsfonts}
\usepackage{amssymb}
\usepackage{amsbsy}
\usepackage{hyperref}
\usepackage{setspace}

\def\>{\rangle}
\def\<{\langle}
\def\be{\begin{equation}}
\def\ee{\end{equation}}
\def\bee{\begin{equation*}}
\def\eee{\end{equation*}}
\def\ft{\footnotesize}
\begin{document}

\noindent
\footnotesize{

\Large{\textbf{Vibration-enhanced energy transfer in living molecules}}

\normalsize{Vlatko Vedral\footnote{Centre for Quantum Technologies, Block S15, 3 Science Drive 2, National University of Singapore, Singapore 117543}$^{,}$\footnote{Department of Atomic and Laser Physics, Clarendon Laboratory, University of Oxford, Parks Road, Oxford OX1 3PU, U.K. $^{*}$email: vlatko.vedral@qubit.org}$^{*}$ and Tristan Farrow$^{1}$ }

\textbf{
The conversion of an absorbed photon from the exciton energy into the reaction centre in the photosynthetic complex has a near unit efficiency. It is becoming clear that any classical model, where the exciton hopping is modeled by a classical stochastic diffusion equation, cannot explain such a high degree of efficiency.  A number of different quantum models have been proposed, ranging from a purely unitary model with long range exciton interactions to a noise-aided stochastic resonance models. Here we propose a very simple spin-boson model that captures all the features of the efficient part of energy transfer. We show how this model describes a scenario where a donor-acceptor system can be brought into resonance by a narrow band of vibrational modes so that the excitation transfer between the two can be made arbitrarily high. This is then extended to a seven exciton system such as the widely-studied FMO photosynthetic complex to show that a high efficiency is also achievable therein. Our model encodes a number of readily testable predictions and we discuss its generalisations to include the localisation in the reaction centre.
}

\textbf{Introduction}

The cofounders of quantum theory, Werner Heisenberg and Erwin Schr\"odinger, predicted that an entirely new picture would emerge when quantum theory would be applied within the realm of biology~\cite{schrod, heis}. ``Most biologists today still use the language and the way of thinking of classical mechanics; that is, they describe their molecules as if the parts of the molecules were just stones or something like that [...] but I feel that sooner or later, also in biology, one will come to realize that this simple use of pictures, models, and so on will not be quite correct."~\cite{heis} We are presently at a juncture when we can credibly speak of entering an era of 'quantum biology'~\cite{vlatko09}, designating the gathering critical mass of surprising experimental results~\cite{fleming04, fleming05, fleming07sci, col10, engel10} that support Heisenberg's and Schr\"odinger's predictions more than half a century later. Spectroscopists are observing wavelike electronic quantum coherences in the photosynthetic molecules~\cite{fleming04, fleming05, fleming07sci, col10, engel10}. The light-harvesting proteins absorb light and transport the light-energy to a reaction centre with near-perfect efficiency, unmatched by any artificial system.

We adopted the principle of 'minimum design' to develop a simple model which explains how energy can be transported with near perfect efficiency from a captured photon to the reaction centre of a typical light-harvesting protein. Considering that primitive photosynthetic cells (Stromatolites) appeared over three billion years prior to any other more complex life-forms, it is not illogical to assume that nature has designed a photosynthetic mechanism using minimal resources but honed to near-perfection under the action of evolution. It is also worth bearing in mind that the eruption of land-based plant life around 400 million years is a relatively recent phenomenon in the evolutionary history of photosynthesis. In this perspective, the extreme efficiency of the photosynthetic energy-transfer process is perhaps less surprising.

First, we need a discussion to justify the use of coherent quantum behaviour in complex bio-molecules. In biological processes we typically have four different relevant timescales, each separated by about three orders of magnitude. The fastest are the optical excitation timescales, which are typically of the femtosecond duration. The next would be the timescales of coherent hopping of excitons which are typically on the order of picoseconds. This is followed by chemical reactions that convert excitons into something like the ATP cycle and they are typically on the order of microseconds. Finally, there are longer, macroscopic, timescales on which things like neurological response take place and this is milliseconds and longer (up to a second).

From a simple heuristic argument, it is immediately clear that coherent quantum effects can realistically only be present within the first two time-scales. Take a typical molecule of roughly hundred thousand atoms (smaller molecules than that are not really likely to play a meaningful biological role). Imagine that a photon is absorbed which puts this molecule into a conformal superposition of different spatially arranged states. Such conformal changes are ubiquitous in living molecules.  How long will this superposition survive? A simple analysis suggests that the ratio of decoherence time to dissipation time is $\hbar^2 / (mx^2 kT)$, where \emph{m} is the mass of the molecule, \emph{x}, is the coherence length and is a few nanometers (comparable to the molecular size), and \emph{T}, the temperature, $\sim 300~K$. The dissipation, then, is typically of the order of seconds to milliseconds, which in turn leads to decoherence times of the of the order of nano- to picoseconds. A more detailed and rigorous analysis~\cite{leg} leads to the same conclusion. Thus, only the first two biological stages can survive the environmental influences and be quantum-coherent over and above what might be expected of processes that do not survive long enough to affect the dynamics. These two timescales, the femto- and picosecond domains, are exactly the ones relevant for the efficiency of photosynthesis we are attempting to explain.

\begin{figure}[htbp]
\centering
\includegraphics[angle=0, width=6cm]{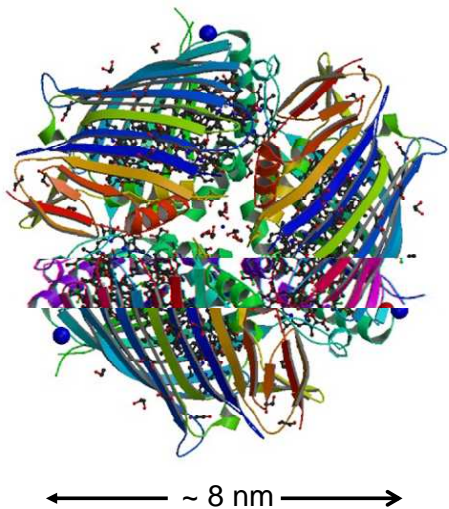}
\caption{\ft{Structure of the light-harvesting pigment protein enabling excitonic energy transfer in the green sulphur photosynthetic bacteria $Prosthecochloris$ $aestuarii$~\cite{blanken}.}}
\label{fmo}
\end{figure}

A widely-used tool for calculating energy transfer rates between molecules is given by F\"orster theory~\cite{forster}. This yields a rate for the energy transport from the overlap between the absorption and emission spectra of donor and acceptor states. In the semi-classical picture, as presented by F\"orster theory, energy is transferred by the incoherent hopping of energy excitations (excitons) down an energy ladder, from the light-harvesting protein to the reaction centre. The nanoscale dimensions and dense packing of the pigment proteins, where intermolecular separations are similar to or smaller than the size of the molecules, however, lets us postulate that the protein environments can be strongly correlated. The picture that emerges, then, is quantum mechanical: excitons retain their wavelike character and exist in a superposition of quantum states. This allows them to sample the different energy pathways and to select the fastest route to the reaction centre almost instantaneously and at no energy cost. Such is the picture borne out by the recent experiments~\cite{fleming04, fleming05, fleming07sci, col10, engel10}. F\"orster theory does not foresee coherences between the donor and acceptor states. However, without those it is impossible to adequately explain the rapid transfer of energy or to account for the near-perfect efficiency of the process. Indeed only few photosynthetic complexes have been adequately characterized by F\"orster theory, suggesting that quantum effects might have a role in facilitating energy-transfer in photosynthesis.

The photosynthetic molecule of FMO bacteriochlorophyll consists of seven holding sites each able to hold an exciton (electron-hole pair)~\cite{scholes06}. When a photon is absorbed by the FMO complex, an electron-hole pair is photoexcited in one of the seven sites. In F\"orster theory, where the exciton hops from site to site until it finds the reaction center where it deposits its energy, fails to account for the hundreds of femtosecond timescales~\cite{fleming07nat, vlatko09} on which the exciton finds its way to the reaction centre. Experiments~\cite{fleming04, fleming05, fleming07sci, col10, engel10} and numerical simulations~\cite{ishi} suggest a coherent superposition of a single exciton state over the seven sites of the FMO complex. From a quantum physical perspective, upon absorption of a photon whose wavelength is larger than the length-scale of the FMO molecule ($\sim~8$nm), the excited state thus created is assumed to take the form of a quantum superposition of excitons. The evolution of the exciton to the reaction centre was previously assumed to be a hopping motion from site to site down an energy ladder until capture at the reaction centre.

Let us first set the context before considering the details of our model. Existing models~\cite{castro08, plenio1, lloyd09} resort to stochastic theory and introduce Markovian statistics, thereby introducing the complexity we are happy to dispense with. We adopt an effectively nonmarkovian analysis and strip out all complexity to produce a model that is as simple as possible, but no simpler. The nonmarkovian nature was also explored in~\cite{ishi}, but contains more complex features that are not needed in our model. Furthermore, the efficiency predicted in theory~\cite{lloyd09} is lower than experimentally measured values, while its trend with increasing temperature diverges from what might be expected from a system designed to protect coherence. Coherent random walk models~\cite{castro08, plenio1, lloyd09} approach the problem of energy transfer with the view that the probability of transfer depends on the exciton being ejected from energy minima in which it comes unstuck, by coupling to a thermal bath. The trapped excitonic states, or dark states, then rely on a coherent random walk to reach the site of the reaction centre. This contrasts with the simple model we present here, which dispenses with the need to introduce noise.

A system that seeks to preserve coherence would, it seems likely, also seek to exclude coupling to a thermal bath. Our starting point, then, picks up where the authors of~\cite{fleming07nat} conclude: ``The FMO light-harvesting complex provides an opportunity to apply more complete energy transfer theories that invoke nonmarkovian dynamics and include coherence transfer. Such theories need to include wavelike energy motion owing to long-lived coherence terms...Further, the observed preservation of coherence in this photosynthetic system requires us to redefine our description of the role of electron-phonon interactions within photosynthetic proteins. In particular, the protein may not only enforce the structure that gives rise to the couplings, but also modulate those couplings with motions of charged residues and changing local dielectric environments, which will change exciton energies and promote coherence transfer."~\cite{fleming07nat}. It is therefore natural that the environment of the FMO complex is not the standard noisy one, but can be thought of as adding extra coherence to the excitons.

The nonmarkovian property of the system and phonon environment dynamics can formally be phrased as follows. Any Markovian dynamics has the property that the distinguishability between input states cannot increase (and usually decreases) with time. If, therefore, we record the opposite trend in the evolution of our system, namely that the distinguishability of states increases, the immediate conclusion is that the system-to-environment coupling is nonmarkovian. The increase in distinguishability is therefore sufficient to witness a nonmarkovian character.
Our model, with two qubits coupled to the narrow-band phonon mode (to be presented below), is nonmarkovian in the sense that the ability to distinguish between different two qubit states can increase with time. This means that information flows coherently from the environment back into the system, contrary to Markovian processes. However, in our semiclassical approximation, where the phonon mode is treated as a static classical field, the ensuing evolution of the two qubits is unitary and there are then of course nonmarkovian effects.

\textbf{Two-level system}

\begin{figure}[htbp]
\centering
\includegraphics[angle=0, width=12cm]{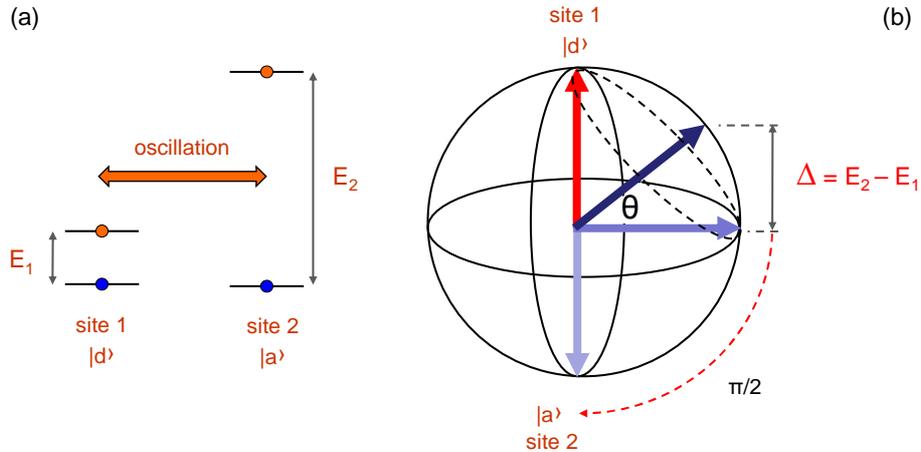}
\caption{\ft{(a) Vibrationally-assisted transition between two sites, 1 and 2, representing the exciton~\cite{scholes06} donor and acceptor states, $|d\>$ and $|a\>$, respectively.  with an energy mismatch. (b) Bloch sphere representation of the transition, illustrating the detuning from resonance between the two states in the two sites. On transition between the two sites, the Bloch vector flips between the two sites which are detuned by an energy deficiency $\Delta$. The parameters are: $\Delta$ is the `coherence restitution' term, $\Delta = E_2 - E_1$, and $E_2 - E_1$ is the energy difference between the excited states of chromophore sites $1$ and $2$ on the photosynthetic molecule. On resonance, $\Delta = g |\alpha|$, which is analogous to Rabi oscillations in atoms.}}
\label{bloch}
\end{figure}

The first model we consider is a simple donor-acceptor one with the F\"orster dipole-dipole exchange interaction. The problem we need to address is the efficiency of excitation transfer from the donor state, $|d\>$, to the acceptor, $|a\>$, given that the two are generally far detuned in energy, as illustrated~\ref{bloch}(a) by the two energy-detuned sites 1 and 2. More precisely, the detuning $\Delta$ is typically of the order of optical frequencies, while the hopping interaction strength $J$ due to the F\"orster coupling is some two orders of magnitude smaller. If there was no other mechanism present, then the excitation would have a very small chance of being faithfully transferred from the donor to the acceptor. An easy way to see this is to focus only on the relevant subspace which leads to an effectively two level system. One level designates the donor energy and the other one the acceptor energy and they are separated by the detuning $\Delta$, while driven at the rate $J$. This is the same as the Rabi model of a two level system driven by an external field. The Bloch sphere representation in figure~\ref{bloch}(b) illustrates the $\theta= \pi/2$ rotation of the Bloch state-vector (indicated by the rotating arrows). This rotation, hence the evolution of system from the donor site $|d \>$ to the acceptor site $|a\>$, is characterised by the well known probability for transfer:
\be
p= \frac{J^2}{J^2+\Delta^2}
\label{proba}
\ee

Given that $J$ is roughly hundred times smaller than $\Delta$ this leads to a probability as small as $10^{-4}$. Clearly when we deal with the near unit efficiencies of energy transfer in photosynthesis, this simple model is insufficient. In practice, however, the donor-acceptor system is part of a complex molecule that has a multitude of vibrational degrees of freedom in addition to the electronic component. There are usually acoustic vibrations (on the order of megahertz) and optical vibrations (roughly three orders of magnitude larger and higher, up to optical frequencies). Given that the transfer is meant to take place at the rate of inverse $J$ (on times scales of picoseconds), it is natural to think that acoustic vibrations will be irrelevant, while the optical ones might play a useful role. Here we present one such simple model where the initially detuned donor and acceptor are shifted into resonance by a narrow band of phonons that are coherently excited.

Typically phonons would couple to excitons in an off-resonant, dispersive, fashion. We are used to thinking that this will naturally lead to the dephasing of excitons, however, this is only really true if, speaking somewhat loosely, the phonons have a broad spectrum and are subject to a weak coupling to excitons. This regime leads to the validity of the Born-Markov approximation~\cite{gard}. If instead we have a strongly coupled narrow band of phonons, then the vibrational degrees of freedom act coherently and can in fact aid the transfer. This is the key mechanism we are proposing in this paper. To see how phonons can enhance the exciton transfer, let us introduce a simplified model of vibrations which will suffice for our purposes.

First we start with the full Hamiltonian, including donor and acceptor, phonons and the electron-phonon coupling:
\begin{equation}
H = \frac{E_1}{2} \sigma_d^z + \frac{E_2}{2} \sigma_a^z + J (\sigma_d^x\sigma_a^x+\sigma_d^y\sigma_a^y) + \sum_n \hbar \omega_n a^{\dagger}_n a_n + \sum_i (g^1_i \sigma^z_1 +g^2_i \sigma^z_2)(a_i+a_i^{\dagger})
\end{equation}
The energy difference between donor and acceptor is $E_2-E_1 = \Delta$, the exchange coupling is $J$, the third term is the phonon energy and the last is the electron-phonon coupling. Now we make the following set of assumptions (which basically define our model):
\begin{enumerate}
\item The phonon density is narrow, so that we can effectively approximate relevant vibrations by a single phonon mode.
\item	That donor and acceptor are exactly out of phase as far as their coupling to vibrations.
\item	 That the vibrational mode can be treated effectively classically.  This means that they are in a coherent state (or near coherent) of some aplitude $\alpha$.
\item	That the frequency of phonons is much smaller than J (to be quantified later).
\end{enumerate}
The Hamiltonian representing the donor-acceptor qubit coupled to a single vibrational mode (assumption 1) is given by:
\begin{equation}
H = \frac{\Delta}{2} (|d\rangle\langle d| - |a\rangle\langle a|)  + J (|d\rangle\langle a| - |d\rangle\langle a|) + \hbar \omega a^\dagger a + (g_1\sigma^z_1 +g_2 \sigma^z_2)(a+a^\dag)
\label{ham}
\end{equation}
Here $\omega$ is the phonon frequency and $g_d$ and $g_a$ are the exciton-phonon coupling strengths for the donor and acceptor respectively. Assumption 2 states that the coupling is of opposite phase which means that we put $g_d=-g_a=g$. This is of central importance to our model since this means that the energy shift acts in such a way as to bring the donor and acceptor into resonance, providing that the product of the phonon coupling strength and its amplitude of oscillation is equal to the detuning. Implementing the assumption 3 effectively means writing the last term as $g \alpha (\sigma^z_d - \sigma^z_a)$, where $\alpha$ is the real part of the amplitude of the phononic coherent state. Because we assume that phonons evolve slowly compared with relevant time scales (assumption 4), the amplitude is for all practical purpose time-independent. The resulting Hamiltonian now effectively represents an DC Stark shift in excitons induced by phonons:
\be
H = \begin{matrix}
\phantom{....} \<d| \qquad \qquad \<a|\\
\begin{matrix}
|d\> \\ |a\>
\end{matrix}
\begin{bmatrix}
\Delta-g |\alpha| & 2J \\
2J & -\Delta+g |\alpha| \\
\end{bmatrix}
\end{matrix}
\ee

If $\Delta = + g\alpha$, the donor-acceptor system is DC Stark shifted into full resonance. This could be the modulation postulated in the cited statement from~\cite{fleming07nat}, namely that the protein may modulate the electron-phonon coupling . Now the transfer due to the ensuing Rabi oscillations becomes on resonant leading to a unit probability of transfer. In reality, of course, the probability will be less than unit. Let us therefore discuss in detail how accurately each of the assumptions has to hold for us to still have a high fidelity of excitation transfer.

There are four main types of errors introduced in our model, all stemming from the approximations introduced in the vibrational degrees of freedom. One is that the phonons will most certainly not be single mode, but will instead have some spread, which we label $\delta \omega$. Each of these modes will couple with a different strength to excitons, say $g_i$, where $i=1,2,...,7$ is the FMO protein site index. The second error could come from the fact that the donor and acceptor are not exactly out of phase as far as the phonon mode. The third type of error comes from the spread in phonon number in each mode. We require phonon states which are number-squeezed, in the sense that the dispersion in phonon numbers is not larger than the root of the mean number of phonons. This is why we use coherent states in our calculations, though the phase coherence is by no means required and the corresponding number-state mixture would also suffice to achieve the same evolution. Finally, the last type of error will come from the fact that we assumed the mode to be static, i.e. that its oscillation frequency is much smaller than $J$. In effect this means approximating $\cos \omega t$ to unity. The error is therefore to the second order equal to $\omega^2 t^2$ which can be estimated to be contributing $g \alpha \omega^2/J^2$ to the detuning between the excitons. Rather than analyse each of errors in more detail, we can make a simple generic statement. It can readily be seen from equation~(\ref{proba}), that for an efficiency decrease of say 5$\%$, the error on $\delta \omega$ (including all possible source) has to be as high as 20$\%$ of the oscillation frequency, since $J \sim 10^{13}$Hz and $p \sim 1 - (\delta \omega / J)^2$, where $\delta \omega$ is the error on $\delta \omega$. This still allows ample overhead for a extremely efficient energy transfer since we can tolerate up to $20$ percent of $J$ inaccuracy in our mechanism to create on resonance.

The central assumption which underpins our model, that the FMO protein ensures coherence enforcing a narrow phononic band, can be tested in experiments aiming to reveal the temperature-invariance of the phononic mode. Although the latest experimental results do not seek to ascertain the profile of the phonon modes as a function of temperature, they confirm that coherence effects persist even at ambient temperature~\cite{col10}. This supports the view that the FMO protein has evolved a specialised mechanism to modulate the electron-phonon interaction~\cite{fleming07nat}. In our model, temperature plays effectively no role as at room temperature we have some $10$ thermal phonons excited in our mode, while the coherent state amplitude is well above and dominates the overall dynamics.

It now becomes appropriate to comment on how nature could create a coherent phonon state? The possible existence of phononic coherent states in living molecules has been discussed for a long time, starting by speculations made by Fr\"ohlich~\cite{fro}. The research is still inconclusive in spite of the vast existing literature on the topic. However, we would like to point out that the mechanical energy due to photon pressure is certainly large enough to provide sufficient number of phonons needed for our mechanics. It can indeed easily be calculated that the average number of photons from the solar flux on earth is about $100$ Watts per meter squared. A typical FMO complex (we assume its area to be tens of nanometers squared) would therefore receive $10^{-28}$ Joules per picosecond. If all of this energy is converted into the mechanical energy of the phonon mode we assumed above this would lead to $\alpha \approx 10^2$ possible during the assumed coherence timescales of the exciton dynamics. The electron-phonon coupling strength required for our mechanism would then be on the order of $g\approx J$. This is consistent with detailed analytical studies presented in~\cite{schulten}. It is, of course, completely outside the scope of the present work to study if this optomechanical conversion of energy of this type is possible in nature.

\textbf{Seven-site system}

Our model has can be widely applied to any coherent energy transfer in biological system. By way of example, here we choose to apply it to
the seven-site system found in photosynthetic molecules. This is convenient because Hamiltonians of many FMO complexes have been much studied and their energies are extensively catalogued. The Hamiltonian we use is represented by a seven-dimensional matrix~\cite{adolphs06, plenio1}.

Before we do that, we point out that although the exciton energies, $E_i$, at sites $i=1,2,...,7$, and the hopping rates $J_{ij}$ between sites are well characterised, we need to use a range of relevant phonon frequencies and their exciton coupling strengths. Again, we assume a narrow-band of vibrational modes which brings these seven sites into resonance. Even though our model is still simple, it turns out to be difficult to treat it analytically. Here we present the results of our numerical investigations.

\begin{figure}[htbp]
\centering
\includegraphics[angle=0, width=7cm]{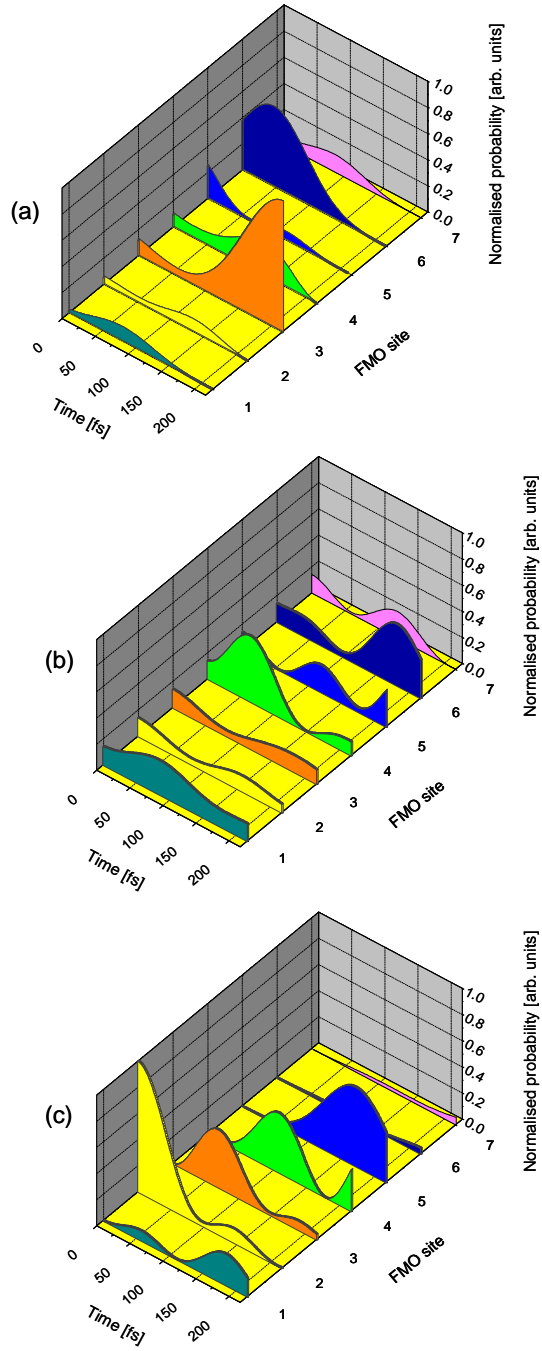}
\caption{\ft{Simulated time-evolution of the probability of finding the exciton in sites 1 to 7 of the FMO complex, initialised with (a) a quantum superposition leading to excitonic confinement in site 3 after $\sim 220$~fs. The evolution shown in (b) is for an initial state in an equal superposition of the exciton in all seven sites, and (c), for an arbitrarily chosen initial pure state in site 2. Plotted here, are only the relevant diagonal elements, which were extracted from the evolution of the full state containing all the off-diagonal elements too.}}
\label{evolution}
\end{figure}

The probability distribution of an exciton in a quantum superposition across the seven sites of an the $Prosthecochloris$ $aestuarii$ light harvesting protein evolves in time. Figure~ref{evolution} (a) shows the evolution of the quantum superposition resulting in the capture of the exciton by site 3 within $\sim~200$~fs with 99.99\% probability. Once the exciton becomes confined in site 3, its energy is deposited in the reaction centre. We find, however, that when the excitonic wave-function is superposed equally over all seven sites, as shown in (b), the probability of channeling confining the exciton in site 3 fails to reach unity, even at multiple periods of the evolution within the timescales of relevant biological processes. A similar outcome occurs in (c), where the system is initialized with an exciton localised in a single site, here chosen arbitrarily as site 2 (since our simulations showed that this was site-independent). Our simulations further indicate that for the $Pr.$ $aestuarii$ complex, the excitonic wavefunction favours site 6, which has a higher occupation probability, $\sim~20$\%. This is to be expected considering that the exciton energies are different and therefore couple with different strengths to sunlight. The existing experimental research is insufficient to decide with more certainty exactly what the initial state of the FMO complex is, but it would certainly be surprising to find all cites excited with the same quantum mechanical amplitude. Although the energetic distribution of seven sites is apparently random and asymmetric, we can speculate that the FMO proteins have evolved to capture photons with an energy that is likely to generate an exciton state with the optimal superposition for energy transfer.

\textbf{Discussion and conclusion}

Finally, although our model explains the near unit efficiency of transfer, it does not seek to explain how the exciton becomes trapped in the reaction centre. That is quite a separate but obviously related problem, since without trapping, the excitation will for ever oscillate between sites under the action of the unitary evolution. It is foreseeable, however, that our model could with a small modification also tackle the problem of the irreversibility of the energy transfer. When the exciton is captured by the site at the reaction centre, a mechanism inspired from the physics of quantum dots can be introduced, where an electron flopping between two potential traps can be confined in one of them when a rapidly oscillating field of large amplitude is applied at the boundary ~\cite{naz}. Whether this mechanism, which is analogous to conventional dephasing, occurs in FMO complexes is clearly open to further investigation, but it is forseeable that the requirement on the timing of dephasing need not be stringent.

The appeal of the model presented here, then, is that it offers a simple mechanism from which we predict the coherence effects observed by experimentalists~\cite{fleming04, fleming05, fleming07sci, col10, engel10}. This is without recourse to Markovian dynamics or quantum entanglement~\cite{whaley09, plenio2}. We adhere instead to the principle of minimum design to explain how energy can be transferred with near unit efficiency to the reaction centre of an FMO complex. Although quantum entanglement may well be a feature photosynthesis, our model suggests that is remains a redundant by-product without clear effect on the efficiency of the excitonic dynamics.

\textbf{Acknowledgements}
The authors acknowledge financial support from the National Research Foundation and Ministry of Education in Singapore as well as the Engineering and Physical Sciences Research Council, the Royal Society and the Wolfson Trust in UK. V.V. is a fellow of Wolfson College, Oxford.

}

\end{document}